\documentclass{PoS}

\title{Studying and removing effects of fixed topology in a quantum mechanical model}

\ShortTitle{Studying and removing effects of fixed topology}

\author{\speaker{Arthur Dromard}\\
        Goethe-Universit\"at Frankfurt am Main, Institut f\"ur Theoretische Physik, \\ $\quad$ Max-von-Laue-Stra{\ss}e 1, D-60438 Frankfurt am Main, Germany\\
        E-mail: \email{dromard@th.physik.uni-frankfurt.de}}

\author{Marc Wagner\\
        Goethe-Universit\"at Frankfurt am Main, Institut f\"ur Theoretische Physik, \\ $\quad$ Max-von-Laue-Stra{\ss}e 1, D-60438 Frankfurt am Main, Germany\\
        E-mail: \email{mwagner@th.physik.uni-frankfurt.de}}

\abstract{
At small lattice spacing, or when using e.g.\ overlap fermions, lattice QCD simulations tend to become stuck in a single topological sector. Physical observables then differ from their full QCD counterparts by $1/V$ corrections, where $V$ is the spacetime volume. Brower et al.\ and Aoki et al.\ have derived equations by means of a saddle point approximation, to determine and to remove these corrections. We extend these equations and apply them to a simple toy model, a quantum mechanical particle on a circle in a square well potential at fixed topology. This model can be solved numerically up to arbitrary precision and allows to explore effects arising due to fixed topology. We investigate the range of validity and accuracy of the above mentioned equations, to remove such fixed topology effects.
}

\newcommand{\ltapprox}{\raisebox{-0.5ex}{$\,\stackrel{<}{\scriptstyle\sim}\,$}}

\FullConference{31st International Symposium on Lattice Field Theory - LATTICE 2013\\
		July 29 - August 3, 2013\\
		Mainz, Germany}

\begin{document}


\section{Introduction }

Topology freezing or fixing are important issues in quantum field theory, in particular in QCD. 
For example, when simulating chirally symmetric overlap quarks, the corresponding algorithms do not allow transitions between different topological sectors, 
i.e.\ they fix the topological charge (cf.\ e.g.\ \cite{Aoki:2008tq}).
Also when using other quark discretizations, e.g.\ Wilson fermions, topology freezing is expected at lattice spacings 
$a \ltapprox 0.05 \, \textrm{fm}$, which are nowadays still rather fine, but realistic \cite{Luscher:2011kk}. 
There are also applications, where one might fix topology on purpose. 
For example, when using a mixed action setup with light overlap valence and Wilson sea quarks, 
approximate zero modes in the valence sector are not compensated by the sea. The consequence is an ill-behaved continuum limit \cite{Cichy:2010ta}.
Since such approximate zero modes only arise at non-vanishing topological charge, fixing topology to zero might be a way to circumvent the problem.
A possible solution to this problem is to restrict computations to a
single topological sector, either by sorting the generated gauge link
configurations with respect to their topological charge or by directly
employing so-called topology fixing actions (cf.\ e.g.\
\cite{Fukaya:2005cw,Bietenholz:2005rd,Bruckmann:2009cv}).

In view of these issues it is important to develop methods,
which allow to obtain physically meaningful results (i.e.\ results corresponding to unfixed topology) from fixed topology simulations. 
The starting point for our work are the seminal papers \cite{Brower:2003yx,Aoki:2007ka}. 
We extend the calculations from these papers and propose and test a corresponding method by applying it to a simple quantum mechanical toy model. 
Similar recent investigations using the Schwinger model can be found in \cite{Bietenholz:2011ey,Bietenholz:2012sh,Czaban:2013}.


\section{Working at fixed topology}


\subsection{Hadron masses at unfixed and at fixed topology and their relation}

The partition function and the temporal correlation function of a hadron creation operator $O$ at fixed topological charge $Q$ and finite spacetime volume $V$ are given by
\begin{eqnarray}
\label{eq:ZQ} & & \hspace{-0.7cm} Z_{Q,V} \ \ = \ \ \int DA \, D\psi \, D\bar{\psi}\, \delta_{Q,Q[A]} e^{-S_E[A,\bar{\psi},\psi]} \\
\label{eq:CQ} & & \hspace{-0.7cm} C_{Q,V}(t) \ \ = \ \ \frac{1}{Z_{Q,V}} \int DA \, D\psi \, D\bar{\psi} \, \delta_{Q,Q[A]} O(t) O(0) e^{-S_E[A,\bar{\psi},\psi]} .
\end{eqnarray}
For sufficiently large $V$ one can use a saddle point approximation and expand the correlation function according to
\begin{equation}
\label{eq:CQ_} C_{Q,V}(t) \ \ = \ \ A_{Q,V} e^{-M_{Q,V} t}  \quad,\quad M_{Q,V} \ \ = \ \ M(0) + \frac{M''(0)}{2 \chi_t V} \bigg(1 - \frac{Q^2}{\chi_t V}\bigg) + \mathcal{O}(1/V^2)
\end{equation}
%
\cite{Brower:2003yx}, where the expansion is in the three parameters $M''(0) t / \chi_t V$, $1 / \chi_t V$ and $Q^2 / \chi_t V$. $M_{Q,V}$ is the mass of the hadron excited by $O$ at fixed topological charge $Q$ and finite spacetime volume $V$, $M(\theta)$ is the hadron mass in a $\theta$-vacuum at infinite $V$ (cf.\ e.g.\ \cite{Coleman:1978ae}), $M(0) = M(\theta = 0)$ is the physical hadron mass (i.e.\ the hadron mass at unfixed topology), $M''(0) = d^2M(\theta) / d\theta^2|_{\theta = 0}$ and $\chi_t$ denotes the topological susceptibility.


\subsection{\label{sec:method}Determining hadron masses at unfixed topology from fixed topology simulations}

A straightforward method to determine physical hadron masses (i.e.\ hadron masses at unfixed topology) from fixed topology simulations based on the equations of the previous subsection has been proposed in \cite{Brower:2003yx}:
\begin{enumerate}
\vspace{-0.3cm}
\item Perform simulations at fixed topology for different topological charges $Q$ and spacetime volumes $V$, for which the expansion (\ref{eq:CQ_})  is a good approximation, i.e.\ where $M''(0) t / \chi_t V$, $1 / \chi_t V$ and $Q^2 / \chi_t V$ are sufficiently small. Determine $M_{Q,V}$ using (\ref{eq:CQ_}) for each simulation.

\vspace{-0.3cm}
\item Determine the physical hadron mass $M(0)$ (the hadron mass at unfixed topology and infinite spacetime volume), $M''(0)$ and $\chi_t$ by fitting (\ref{eq:CQ_}) to the masses $M_{Q,V}$ obtained in step~1.
\end{enumerate}


\subsection{\label{sec:imp_exp}Improving the expansion of the correlation function}

In the following we improve the expansions (\ref{eq:CQ_}) by explicitly calculating higher orders proportional to $1/V^2$ and $1/V^3$. The starting point for this calculus has been a general discussion of these higher orders for arbitrary $n$-point functions at fixed topology \cite{Aoki:2007ka}. The lengthy result, which we will derive in detail in an upcoming publication, is
\begin{eqnarray}
\nonumber & & \hspace{-0.8cm} C_{Q,V}(t) \ \ = \ \ \frac{A}{(1+x_2)^{1/2}} \exp\bigg\{-M(0) t \\
\nonumber & & \hspace{-0.6cm} + \frac{1}{\chi_t V} \bigg[-\frac{1}{8} \frac{\mathcal{E}_4}{\chi_t} \bigg(\frac{1+x_4}{(1+x_2)^2}-1\bigg) + \frac{Q^2}{2} \frac{x_2}{(1+x_2)}\bigg] \\
\nonumber & & \hspace{-0.6cm} + \frac{1}{(\chi_t V)^2} \bigg[+\frac{1}{12} \frac{\mathcal{E}_4^2}{\chi_t^2} \bigg(\frac{(1+x_4)^2}{(1+x_2)^4}-1\bigg) - \frac{1}{48} \frac{\mathcal{E}_6}{\chi_t} \bigg(\frac{1+x_6}{(1+x_2)^3}-1\bigg) +\frac{Q^2}{4} \frac{\mathcal{E}_4}{\chi_t} \bigg(\frac{1+x_4}{(1+x_2)^3}-1\bigg)\bigg]\\
%
%
\nonumber & & \hspace{-0.6cm} + \frac{1}{(\chi_t V)^3} \bigg[-\frac{5021}{9216} \frac{\mathcal{E}_4^3}{\chi_t^3} \bigg(\frac{(1+x_4)^3}{(1+x_2)^6}-1\bigg) - \frac{1}{384} \frac{\mathcal{E}_8}{\chi_t} \bigg(\frac{1+x_8}{(1+x_2)^4}-1\bigg) \\
\nonumber & & \hspace{0.4cm} +\frac{19}{768} \frac{\mathcal{E}_4 \mathcal{E}_6}{\chi_t^2} \bigg(\frac{(1+x_4) (1+x_6)}{(1+x_2)^5}-1\bigg) - \frac{Q^2}{3} \frac{\mathcal{E}_4^2}{\chi_t^2} \bigg(\frac{(1+x_4)^2}{(1+x_2)^5}-1\bigg) \\
\label{eq:MQ1} & & \hspace{0.4cm} +\frac{Q^2}{16} \frac{\mathcal{E}_6}{\chi_t} \bigg(\frac{1+x_6}{(1+x_2)^4}-1\bigg) - \frac{Q^4}{24} \frac{\mathcal{E}_4}{\chi_t} \bigg(\frac{1+x_4}{(1+x_2)^4}-1\bigg)\bigg]\bigg\} + \mathcal{O}(1/V^4) ,
\end{eqnarray}
where $\mathcal{E}_n$ denotes the $n$-th derivative of the energy density of the vacuum with respect to $\theta$ at $\theta = 0$ (one can show $\chi_t = \mathcal{E}_2$) and $x_n \equiv M^{(n)}(0) t / \mathcal{E}_n V$.

In principle this improved expansion can directly be used in the fitting procedure outlined in section~\ref{sec:method}. Note, however, that there are six additional unknown parameters (compared to (\ref{eq:CQ_})),which have to be determined via fitting: $\mathcal{E}_4$, $\mathcal{E}_6$ and $\mathcal{E}_8$, $M^{(4)}(0)$, $M^{(6)}(0)$ and $M^{(8)}(0)$. A compromise between improvement on the one hand and a small number of parameters on the other hand, which seems to work well in practice (cf.\ section~\ref{sec:test}), is to set these six new parameters to zero. Then only $1/V^2$ and $1/V^3$ corrections, which are associated with the old parameters $M(0)$, $M''(0)$ and $\chi_t$ are taken into account.


\section{\label{sec:test}Testing the method in quantum mechanics}


\subsection{A simple quantum mechanical toy model: a particle on a circle}

\begin{figure}[htb]
\begin{center}
\includegraphics[width=6cm]{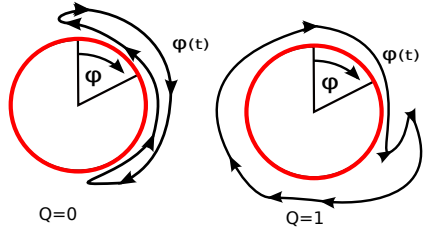}
\caption{\label{FIG001}Classical paths with topological charge $Q=0$ and $Q=1$.}
\end{center}
\end{figure}

To test the method described in section~\ref{sec:method}, we decided for a simple toy model, a quantum mechanical particle on a circle in a square well potential:
\begin{eqnarray}
 & & \hspace{-0.7cm} S_E(\theta) \ \ \equiv \ \ \int_0^V dt \, \bigg(\frac{I}{2} \dot{\varphi}^{2} + U(\varphi)\bigg) - i \theta \underbrace{\frac{1}{2 \pi} \int_0^V dt \, \dot{\varphi}}_{= Q} \\
 & & \hspace{-0.7cm} U(\varphi) \ \ \equiv \ \ \begin{cases} +U_0 & \textrm{if }-(\pi-L) < \varphi < +(\pi-L) \\ 0 & \textrm{otherwise} \end{cases}
\end{eqnarray}
($V$ denotes the finite extent of the periodic time [the analog of the spacetime volume in QCD]). This model shares some characteristic and important features of QCD: the existence of topological charge (paths with topological charge $Q=0$ and $Q=+1$ are sketched in Figure~\ref{FIG001}), the symmetry $+\theta \leftrightarrow -\theta$ and the existence of both bound states and scattering states. Moreover, it can be solved numerically up to arbitrary precision (no simulations required). For the results presented in this section we have used $I = 1$, $U_0 = 10$ and $L = 9 \pi / 10$.

Since parity is a symmetry, the energy eigenstates can be classified according to $P = +$ and $P = -$ (in the following energy eigenvalues of $P = +$ [$P = -$] states are denoted by $E_0^+(\theta)$ [$E_0^-(\theta)$], $n = 0,1,2,\ldots$). The ``mass'' $M(\theta)$ we are going to study in the following is defined as the energy difference between the ground state (which has $P = +$) and the lowest energy eigenstate in the $P = -$ sector, i.e.\ $M(\theta) \equiv E_0^-(\theta) - E_0^+(\theta)$. A suitable creation operator $O$ for a temporal correlation function $C(t)$, whose exponential behavior yields $M(\theta)$, is $O \equiv \sin(\varphi)$.


\subsection{Solving the model and testing the method}

(\ref{eq:CQ_}) as well as (\ref{eq:MQ1}) are expansions in the small parameters $M''(0) t / \chi_t V$, $1 / \chi_t V$ and $Q^2 / \chi_t V$. We are interested to estimate upper bounds for these parameters such that the determination of $M(0)$ as outlined in section~\ref{sec:method} is sufficiently precise. We proceeded as follows:

\begin{enumerate}

\item Solve Schr\"odinger's equation with  $\;H(\theta) \ = \  \left(p - \theta/2 \pi\right)^2/2I + U(\varphi)$.
%
%
Use the resulting energy eigenvalues $E_n^+(\theta)$ and $E_n^-(\theta)$ to determine $M(0)$, $M^{(2)}(0)$, $M^{(4)}(0)$, $M^{(6)}(0)$, $M^{(8)}(0)$, $\chi_t$, $\mathcal{E}_4$, $\mathcal{E}_6$ and $\mathcal{E}_8$, the parameters of the $C_{Q,V}$ expansions (\ref{eq:CQ_}) and (\ref{eq:MQ1}).

\item Calculate $C_{\theta,V}(t)$ using the energy eigenvalues from step~1 and the corresponding wave functions. Perform a Fourier transformation to obtain $C_{Q,V}(t)$, the exact correlation function at fixed topology.
%
%
Define and calculate the effective mass
\begin{equation}
\label{eq:M_ex} M_{Q,V}^{\small \textrm{eff}}(t) \ \ \equiv \ \ -\frac{d}{dt} \ln\Big(C_{Q,V}(t)\Big) .
\end{equation}

\item Determine expansions for the effective mass using (\ref{eq:CQ_}), (\ref{eq:MQ1}) and (\ref{eq:M_ex}) and compare and/or fit the resulting expressions to their exact counterpart (\ref{eq:M_ex}).
\end{enumerate}
Note that in QCD the exact correlator $C_{Q,V}(t)$ and the corresponding exact effective mass $M_{Q,V}^{\scriptsize \textrm{eff}}(t)$ at fixed topological charge $Q$ and spacetime volume $V$ will be provided by lattice simulations.


\subsubsection{The effective mass at fixed topology}

\begin{figure}[htb]
\begin{center}
\includegraphics[width=11.5cm,height=5.5cm]{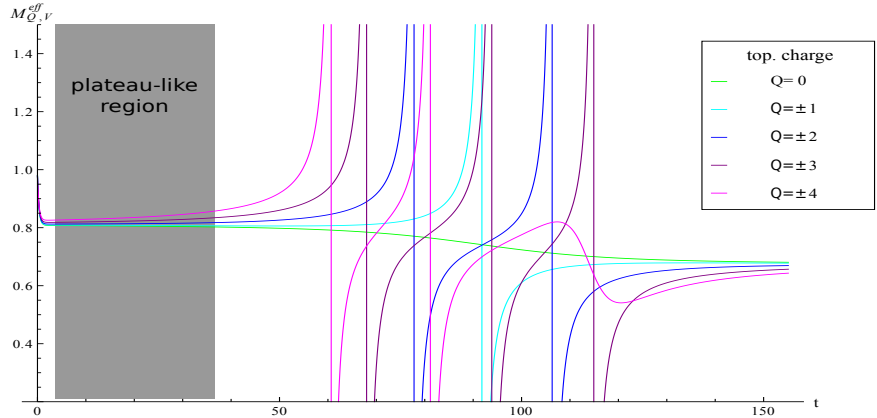}
\caption{\label{fig:M(t)}Effective masses $M_{Q,V}^{\scriptsize \textrm{eff}}$ as functions of the temporal separation $t$ for different topological charges $Q$ and fixed $V = 6 / \chi_t$.}
\end{center}
\end{figure}

In Figure~\ref{fig:M(t)} we show effective masses (\ref{eq:M_ex}) (exact results, not expansions) as a functions of the temporal separation for different topological charges $Q$. As usual at small temporal separations the effective masses are decreasing, due to the presence of excited states. At large temporal separations there are also severe deviations from a constant behavior. This is in contrast ordinary quantum mechanics or quantum field theory (at unfixed topology) and is caused by topology fixing. At intermediate temporal separations there are plateau-like regions (shaded in gray in Figure~\ref{fig:M(t)}), where the expansions (\ref{eq:CQ_}) and (\ref{eq:MQ1}) will turn out to be rather accurate approximations (cf.\ section~\ref{sec:comp_exp}). Note that with increasing topological charges $Q$ the plateau-like regions of $M_{Q,V}^{\scriptsize \textrm{eff}}(t)$ become smaller. A similar trend is observed for decreasing temporal extension $V$.

\subsubsection{\label{sec:comp_exp}Comparing the exact effective mass and the expansions}

\begin{figure}[htb]
\begin{center}
\includegraphics[width=15cm,height=4.8cm]{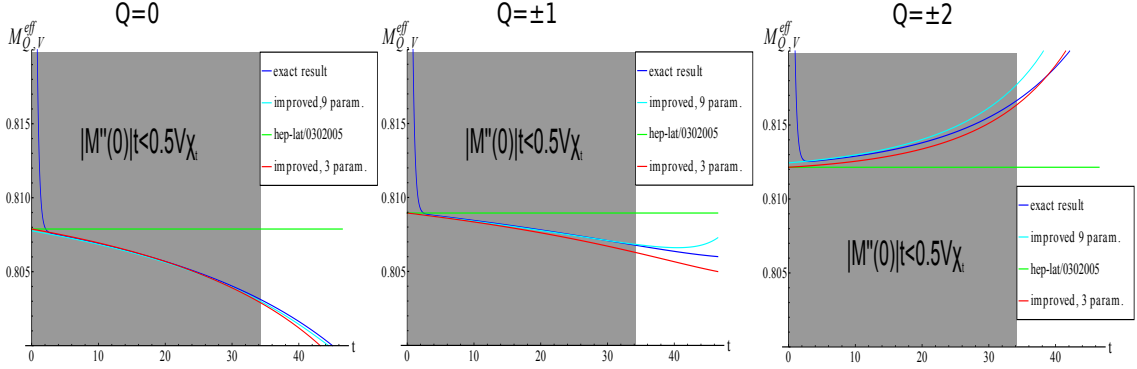}
\caption{\label{fig:Comparison-m}Comparison of the exact effective mass $M_{Q,V}^{\scriptsize \textrm{eff}}$ and various expansions as functions of the temporal separation $t$ for different topological charges $Q$ and fixed $V = 6 / \chi_t$.}
\end{center}
\end{figure}

Figure~\ref{fig:Comparison-m} shows again the gray region of Figure~\ref{fig:M(t)}, where the three panels correspond to $Q=0, \pm 1, \pm 2$. This time not only the exact effective mass is plotted (blue curve), but also various expansions: \\
\phantom{XXX} \textbf{green curve}, the expansion (\ref{eq:CQ_}) from \cite{Brower:2003yx} (three parameters); \\
\phantom{XXX} \textbf{cyan curve}, our improved expansion (\ref{eq:MQ1}) with nine parameters (cf.\ section~\ref{sec:imp_exp}); \\
\phantom{XXX} \textbf{red curve}, our improved expansion (\ref{eq:MQ1}) with three parameters (cf.\ section~\ref{sec:imp_exp}). \\
Clearly the two improved expansions (cyan, red) are much closer to the exact result (blue), than the unimproved expansion (green). Since there does not seem to be a qualitative difference between the two improved expansions, the version with only three parameters (red) seems to be the best candidate for our model to determine the mass $M(0)$ at unfixed topology via fitting.

Figure~\ref{fig:Comparison-m} as well as similar plots for many different topological charges $Q$, 
temporal extensions $V$ and parameters of the model allow to crudely estimate a region, 
where the deviations between our improved expansions (\ref{eq:MQ1}) 
and the exact result (\ref{eq:M_ex}) is $\ltapprox 1 \%$. 
This region corresponds to $|M''(0)| t / \chi_t V \ltapprox 0.5$, $1 / \chi_t V \ltapprox 0.5$ and $Q^2 / \chi_t V \ltapprox 1$.

\subsubsection{Determining the physical mass from fixed topology results}

\begin{figure}[htb]
\begin{center}
\includegraphics[width=12cm]{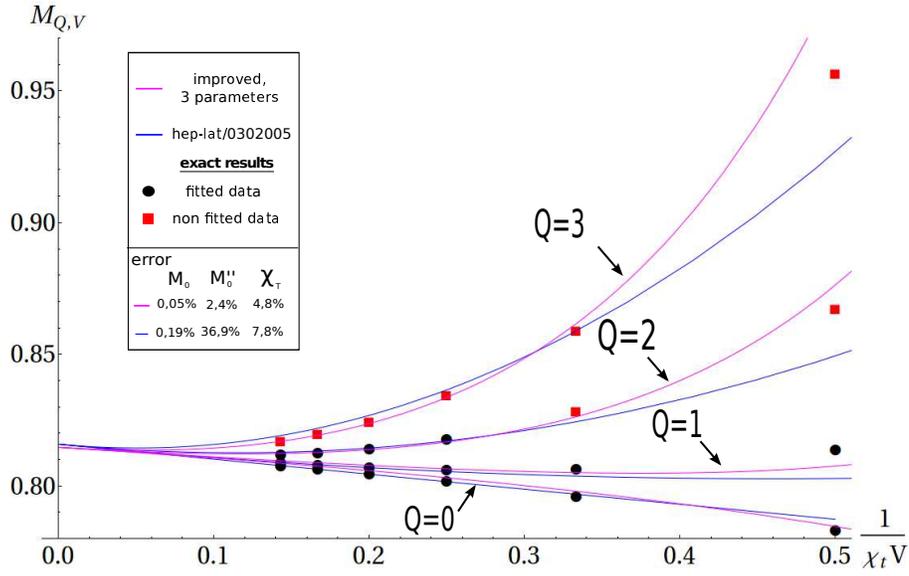}
\caption{\label{fig:Fitting}Determining the physical mass (i.e.\ the mass at unfixed topology) from fixed topology results.}
\end{center}
\end{figure}

In Figure~\ref{fig:Fitting} we mimic the method to determine a physical mass (i.e.\ at unfixed topology) outlined in section~\ref{sec:method}. 
We use the exact result for the effective mass to generate masses at fixed topological charge and finite temporal extent,
$M_{Q,V} \equiv M_{Q,V}^{\scriptsize \textrm{eff}}(t=10)$ (the dots in Figure~\ref{fig:Fitting}, 
step 1 in section~\ref{sec:method}). Then we perform a single fit of either the expansion (\ref{eq:CQ_}) from \cite{Brower:2003yx} 
or our improved version (\ref{eq:MQ1}) with three parameters at $t=10$ inserted in (\ref{eq:M_ex}) to these masses $M_{Q,V}$ 
(only values fulfilling $1 / \chi_t V \ltapprox 0.5$ and $Q^2 / \chi_t V \ltapprox 1$ are taken into account; cf. section~\ref{sec:comp_exp})  , to determine $M(0)$ (the physical mass at unfixed topology), $M''(0)$ and $\chi_t$ (the curves in Figure~\ref{fig:Fitting}, step 2 in section~\ref{sec:method}). Both expansions give rather accurate results for $M(0)$ (the error is of the order of $0.1 \%$) and quite reasonable results for $\chi_t$ (an error of a few percent). Note that the error for both $M(0)$ and $\chi_t$ is significantly smaller, 
when using the improved expansion (\ref{eq:MQ1}), as shown in Figure~\ref{fig:Fitting}.

\section{Conclusion}

We have tested a method to extract physical masses (i.e.\ masses at unfixed topology) from calculations or simulations at fixed topology and finite temporal extent or spacetime volume. The method provides accurate results with errors significantly below $1\%$, when applied to a quantum mechanical toy model. Therefore, it might be a promising candidate to eliminate unwanted fixed topology effects also in QCD.

The method is based on an expansion in $1/V$ (inverse powers of the spacetime volume). We have improved this expansion from \cite{Brower:2003yx} by including higher orders proportional to $1/V^2$ and $1/V^4$ and we demonstrated that these higher orders significantly reduce the associated error.

We also explored the range of validity of the method, 
when applied to our quantum mechanical toy model: $|M''(0)| t / \chi_t V \ltapprox 0.5$, $1 / \chi_t V \ltapprox 0.5$ and
$Q^2 / \chi_t V\ltapprox 1$ corresponds to the above mentioned accuracy of around $0.1\%$.
Assuming that similar conditions hold for QCD, the range of validity of the method should not pose a problem, 
since in typical nowadays QCD simulations $\chi_{t} V = \mathcal{O}(10)$.


\begin{acknowledgments}

We thank  Wolfgang Bietenholz, Krzysztof Cichy, Christopher Czaban, Dennis Dietrich, Gregorio Herdoiza and Karl Jansen for discussions. M.W.\ acknowledges support by the Emmy Noether Programme of the DFG (German Research Foundation), grant WA 3000/1-1. This work was supported in part by the Helmholtz International Center for FAIR within the framework of the LOEWE program launched by the State of Hesse.

\end{acknowledgments}


\end{document}